\documentclass[%
 reprint,
 amsmath,amssymb,
 aps,
]{revtex4-2}

\usepackage{graphicx}% Include figure files
\usepackage{dcolumn}% Align table columns on decimal point
\usepackage{bm}% bold math
\usepackage{xcolor}

\begin{document}

\preprint{APS/123-QED}

\title{Hydrodynamic Collapse of the Leidenfrost Vapor Layer}% Force line breaks with \\

\author{Dana Harvey}
\altaffiliation[Also at ]{Lawrence Livermore National Lab, Livermore, CA 94550, USA}%Lines break automatically or can be forced with \\
\author{Justin C. Burton}%
 \email{justin.c.burton@emory.edu}
\affiliation{%
Department of Physics, Emory University, Atlanta, GA 30033, USA \\
}%

\date{\today}% It is always \today, today,

\begin{abstract}
During the Leidenfrost effect, a stable vapor film can separate a hot solid from an evaporating liquid. Eventually, after formation and upon cooling, the vapor layer cannot be sustained and undergoes a violent collapse evidenced by explosive boiling. Computationally, modeling this instability involves an interplay between hydrodynamics, thermodynamics, rapid evaporation, and length-scales from $\mu$m to cm. Selective assumptions, made to reduce computational costs, have limited most previous studies to steady-state investigations. Here, we combine two-phase laminar flow, heat transfer, and evaporation in a finite-element simulation to examine the failure of Leidenfrost vapor layers during cooling. During periods of quiescence, the geometry of the vapor layer agrees well with steady-state lubrication theory. In the simulations, we report the local temperature of the solid at failure, $T_-$, which provides a lower bound for recent experimental work using the same geometric and material conditions. Surprisingly, we find that inertial forces, which are typically ignored in theoretical treatments of the vapor layer, are responsible for initiating the instability leading to failure. 
%Furthermore, these simulations highlight the importance of inertia, assumed negligible for steady state solutions, in driving vapor layer failure.

%\begin{description}
%\end{description}
\end{abstract}

\maketitle

%\tableofcontents

\section{Introduction \protect\\}

The Leidenfrost vapor layer was first noted by J. G. Leidenfrost in his 1756 treatise on the properties of water. He realized that water would not wet the surface of a sufficiently heated spoon and instead floated on a layer of its own vapor \cite{Leidenfrost}. The significance of this discovery has become important for both fundamental fluid mechanics and industrial applications \cite{Querer}. Previous studies of the Leindenfrost effect have focused on the vapor layer geometry \cite{Burton,Sobac2014,Biance,Snoeijer2009,duchemin2005static,lister2008shape,Chakraborty2022}, spontaneous motion and oscillations of drops \cite{Bouillant2,Cousins2012,Lagubeau2011,Linke2006,Gauthier2019,Ma2,Ma}, drop impact on heated surfaces \cite{tran2012drop,castanet2015drop,riboux_gordillo_2016,YAO1988363,gordillo2022initial}, and ``nano-painting'' through particle deposition \cite{Bain,Elbahri}. The temperature at which the vapor layer forms (or fails) is important in all of these examples, yet this temperature is not well-understood. Influencing factors that determine the Leidenfrost temperature include surface roughness \cite{Kim2,Kim,Kruse,KimNano,Jones2019}, hydrophobicity \cite{Liu,Vakarelski,Vakarelski2,KimNano,jiang2022inhibiting,bourrianne2019cold}, solid thermal properties \cite{Freud,Vakarelski, Vakarelski2,Yagov,Hsu,Sher}, liquid temperature \cite{Jouhara,Freud,Sher,Yagov,Yagov2,KimNano}, solid geometry \cite{Bradfield,Huang,Jouhara,Freud, Vakarelski,Vakarelski2,Sher,Yagov}, and liquid impurities \cite{Huang,Abdalrahman,Hsu,KimNano,moreau2019explosive}. Most recently, studies have revealed that there are two separate Leidenfrost temperatures, a higher temperature for formation ($T_+$), and a lower temperature for failure ($T_-$), which are often separated by over 100$^{\circ}$C \cite{Harvey2021, Zhao, chantelot2021leidenfrost, chantelot2021drop}.

\begin{figure*}
    \centering
    \includegraphics[width= \textwidth]{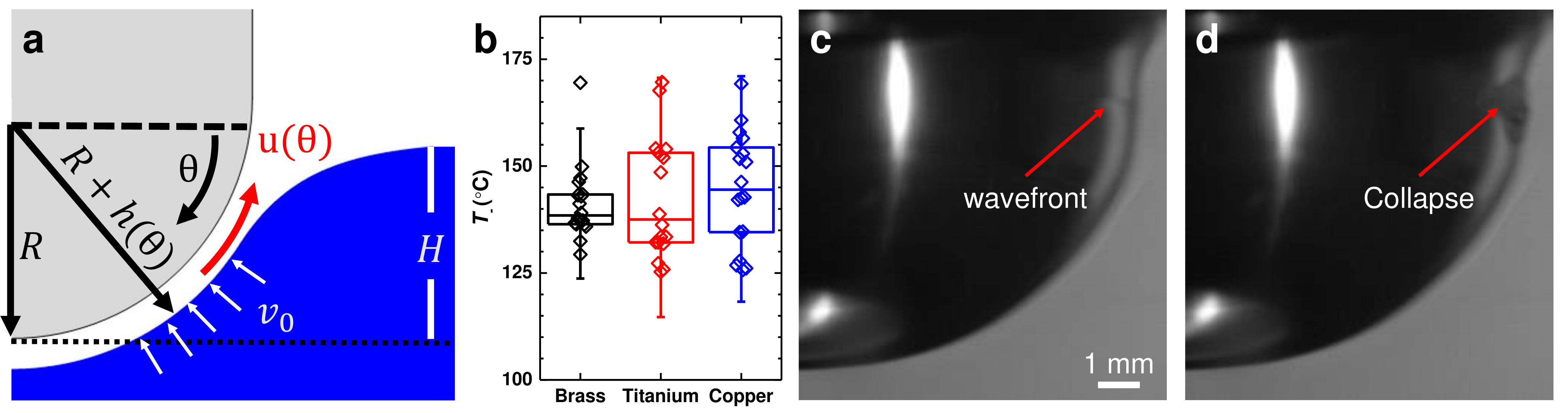}
    \caption{(a) Inverse Leidenfrost schematic. $R$ is the radius of the hot object, $\theta$ is the angular spherical coordinate measured from $\pi/2$ to $\pi$, $h (\theta)$ is the thickness of the vapor layer, $H$ is the height from the tip of the hot object to liquid level, and $v_\text{0}$ and $u(\theta)$ are the evaporation velocity and average flow velocity in the vapor layer. (b) Box and Whisker distribution of experimental values of the failure temperature, $T_-$, as reported in Harvey \textit{et al.} \cite{Harvey2021}. $T_- \approx 140^{\circ}$C for all materials tested. (c) Experimental picture of the wavefront instability \cite{Harvey2021} that leads to the collapse of the vapor layer pictured in (d). The time between the two images is 30.3 $\mu$s.}
    \label{ExperimentDiagram}
\end{figure*}

How does a stable vapor layer fail? On smooth solid materials, $T_+$ is predominantly determined by van der Waals forces between the liquid and solid \cite{Zhao}. Yet prior to failure, Leidenfrost vapor layers can be 10s of microns thick, suggesting that hydrodynamic stability determines $T_-$. In experiments, direct measurements of the gas-fluid and gas-solid interfaces are extremely challenging. Our recent experimental work resolved microsecond dynamics of vapor layer failure using a high speed electrical technique and geometric capacitive model for the fluid interfaces \cite{Harvey2021}. In the experiment, an inverse Leidenfrost system was used, which provided a well-defined vapor layer surface area despite evaporation. The hot solid material was a metallic rod with a spherical tip of radius, $R$ = 8 mm (Fig.\ \ref{ExperimentDiagram}a), immersed in a bath of salt water of height $H$ above the tip. After formation of the vapor layer at high temperatures, the solid was cooled until the vapor layer collapsed and explosive boiling ensued. We found a consistent failure temperature, $T_- \approx 140^{\circ}$C (Fig.\ \ref{ExperimentDiagram}b), for all experiments of various metals and salt concentrations. This suggests that a hydrodynamic instability, \textit{i.e.} surface waves growing in amplitude, incited vapor layer collapse \cite{Harvey2021}. This can be seen in Fig.\ \ref{ExperimentDiagram}c-d and Movie S1. Beyond these experimental results, computational fluid mechanics can be used to specify the dominant hydrodynamic mechanism that initiates vapor layer collapse.

Previous simulations \cite{Snoeijer2009,Sobac2014,Limbeek2017, Chakraborty2022} have succeeded in matching the experimentally-measured geometries of steady-state Leidenfrost drops \cite{Biance, Burton,Limbeek2017}. However, vapor layer failure is inherently dynamic, and steady-state approximations are invalid in this regime. Additionally, other simplifying assumptions commonly used to describe the fluid mechanics of vapor layers may be invalid near failure. First, lubrication theory is typically used to describe the vapor layer, which assumes the thickness (10-100 $\mu$m) is slowly-varying in space and completely ignores inertia \cite{Sobac2014,Chakraborty2022}. The latter is especially important since the Reynolds number in the vapor layer can be of order unity \cite{Ma2}. 
%separating the fluid dynamics of a droplet into a lubrication approximation on the bottom, encompassing the vapor layer (10-100 $\mu$m thick) and the boundary fluid, and a Navier-Stokes fluid on the top with a boundary matching point . A lubrication approximation requires a very small vapor layer thickness compared to the flow length of the evaporating gas. This requirement breaks down locally as a wave peak approaches the hot solid. 
Second, nearly all theory and computational work assumes a uniform temperature throughout the fluid, with few exceptions \cite{Limbeek2017}. A uniform temperature neglects convection, heat transfer, and phase change through evaporation \cite{Chakraborty2022}. These simplifying assumptions significantly reduce the computational costs of modeling vapor layers between solids and liquids. In some cases, different regimes of stability can be found between steady-state solutions \cite{duchemin2005static,lister2008shape,Snoeijer2009}. However, they provide little insight into the rapid dynamics of vapor layer failure.

%heat transfer, phase mixtures, and phase boundary conditions \cite{Chakraborty2022}. Third, time-independence is required for a steady-state solution. However, we know that the dynamics of the vapor must change in time as the system cools for a wave to form. To investigate the instability that causes vapor layer failure, no fluid flows that may lead to surface waves can be neglected. Therefore, a uniform temperature, lubrication approximation, and time-independence cannot be used for this investigation.

Here, we use a COMSOL Multiphysics simulation to combine two-phase laminar flow, heat transfer, and evaporation to investigate the dynamics and failure of Leidenfrost vapor layers upon cooling. The axisymmetric model uses the same inverted Leidenfrost geometry as the experiments, and is validated by matching quiescent interface profiles prior to failure with a steady-state lubrication model, similar to previous studies. Upon cooling, we find that the vapor layer undergoes a rapid and violent failure, similar to experiments. By modeling a wide range of solid material geometries and thermal properties, we demonstrate that the temperature at failure, $T_-$, agrees well with previous experiments, and in fact provides a lower bound on $T_-$ for smooth surfaces. Failure occurs by means of a series of growing sinusoidal perturbations in the liquid-vapor interface that approach the surface. The leading perturbation that touches the surfaces first causes failure at a single point. By varying the surface tension and molecular weight of the gas in the simulation, we found that the inertia of the evaporated gas, almost always neglected in theory, is paramount to inciting the instability which ultimately results in vapor layer collapse.

\section{COMSOL Simulation Details} \label{ComsolDetails}

\begin{figure}
    \centering
    \includegraphics[width=\columnwidth]{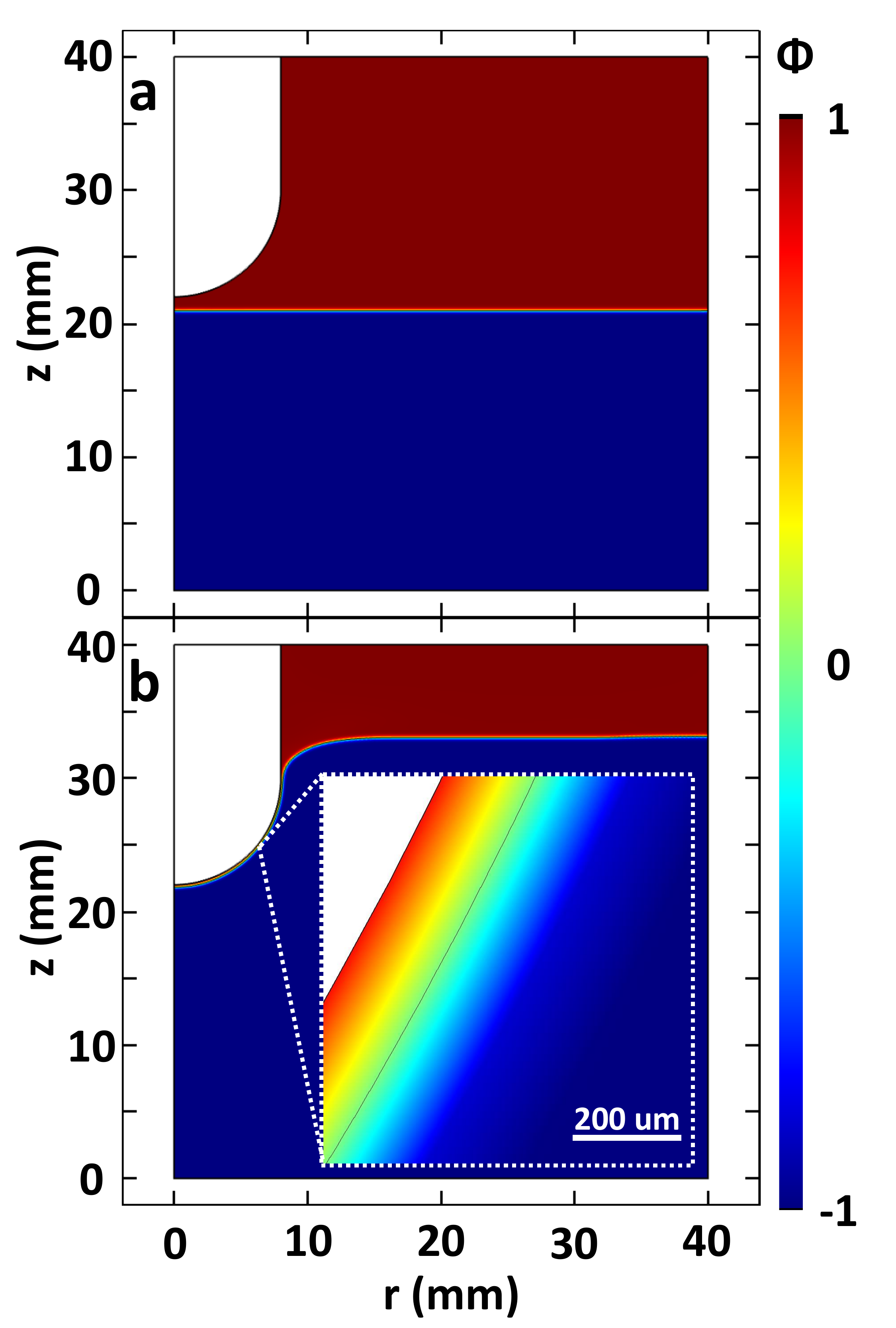}
    \caption{(a) Initial geometry for a hot copper object of $R$ = 8 mm and a water height of $H=14$ mm at $t=0$ s. The color scale denotes the volume fraction of the two-phase fluid (gas = 1, liquid = -1). (b) Characteristic vapor layer profile after $\approx$ 20 s when the water height reached equilibrium. The inset shows a zoom-in of the phase-field profile in the vapor layer. }
    \label{SimulationSize}
\end{figure}

We simulated a three dimensional (3D), axisymmetric inverse Leidenfrost system consisting of a hot object immersed in a heated liquid bath. We used a cylindrical coordinate system with radial coordinate $r$ and axial coordinate $z$. The simulation space was a 40 mm $\times$ 40 mm box consisting of a solid rod with a spherical tip of radius $R$ centered on the axis of symmetry, $r$ = 0 mm, Fig.\ \ref{SimulationSize}a. The bottom tip of the hot solid was fixed at $z$ = 22 mm and the initial liquid height was $z$ = 21 mm. During the simulation, the height of the liquid was slowly raised until it reached a final height $H$, forming a vapor layer around the spherical tip. This process is shown in Movie S2. The surface area of the vapor layer was controlled by varying the radius of the spherical tip, 2 mm $\leq R \leq 16$ mm, and the liquid level 0.5 mm $\leq H \leq 17$ mm, Fig.\ \ref{ExperimentDiagram}a. To suppress superheating of the liquid and minimize wall boundary effects, the outer boundary at $r$ = 40 mm was held at 70 $^{\circ}$C with a slip condition (no restrictions on tangential velocity).

In the fluid domain we modeled a two-phase incompressible fluid with gravity using the non-conservative phase field equations. The non-conservative form must be used to avoid excessive convergence times while maintaining accurate results for our simulations \cite{Sun2007,Chiu2011}. The fluid domain mesh size was adaptive, ranging from 2.25 $\mu$m - 195 $\mu$m. In the solid domain we found a mesh size of 3 $\mu$m - 800 $\mu$m led to reproducible results. Within COMSOL, the laminar flow and two-phase flow modules allow for gradients in density and viscosity, thus the Navier-Stokes equations are defined as follows with the corresponding incompressibility condition \cite{comsol}:
\begin{eqnarray}
    &\rho \dfrac{\partial \vec{\bf v}}{\partial t} + \rho ( \vec{\bf v} \cdot \vec{\nabla})\vec{\bf v}\\
    &= \vec{\nabla} \cdot [-p + \mu(\vec{\nabla} \vec{\bf v} + (\vec{\nabla} \vec{\bf v})^\text{T})] + \rho g\hat{\bf z}, \nonumber\\
    &\rho \vec{\nabla} \cdot \vec{\bf v} = 0.
\end{eqnarray}
Here $\vec{\bf v}$ is the velocity, $g$ is the acceleration due to gravity, $\rho$ is the density of the fluid, $p$ is the pressure, and $\mu$ is the dynamic viscosity of the fluid. 

We note that previous studies investigating steady-state solutions of the vapor layer only consider the fluid dynamics of the vapor flow using a lubrication approximation \cite{Snoeijer2009,Sobac2014}. The lubrication approximation neglects the inertial terms of the Navier-Stokes equations and instead assumes Stokes flow, 
\begin{equation}
   \vec{\nabla} p = \mu \vec{\nabla}^2 \vec{\bf v},
\end{equation}
in the vapor layer, which is typically only 10 $\mu$m - 100 $\mu$m thick and varies slowly along the solid surface. The evaporating liquid is considered a solid boundary with no-slip, or is modeled as a Navier-Stokes fluid  \cite{Sobac2014,Chakraborty2022} where evaporation is parameterized and constant along the interface. We will use this same approximation in Sec.\ \ref{LubeSecs} to verify our full simulation results well before vapor layer failure.

Evaporation was added to the COMSOL model as a heat source term, $Q_s$, defined in the literature as \cite{Jafari2015, DOJO2018}: 
\begin{equation}
    Q_s = -\dot{\text{m}} L \delta,
\end{equation}
\noindent where $L$ is the latent heat of vaporization of the liquid defined in Table\ \ref{Parameters}. The rate of mass vaporization, $\dot{\text{m}}$, and the interface delta function, $\delta$, are defined as:
\begin{equation}
    \dot{\text{m}}= 
    \begin{cases} 
    R_T \rho_v \dfrac{T-T_{sat}}{T_{sat}}, & \text{if } T< T_{sat},\\
    \\
    R_T \rho_l \dfrac{T-T_{sat}}{T_{sat}}, & \text{otherwise},
    \end{cases}
\end{equation}
and
\begin{equation}
    \delta = 3 V_{fl} (1-V_{fl}) \sqrt{|\vec{\nabla}\phi|^2+eps}.
\end{equation}
Here $T_{sat}$, $\rho_v$, and $\rho_l$ are defined in Table\ \ref{Parameters}, $R_T = 0.001$ m/s is a tuning parameter, $eps$ is a small mathematical number to ensure there are no divisions by 0 when using $\delta$ \cite{Jafari2015}, $T$ is the temperature of the fluid, and $V_{fl}$ is the volume fraction of water. In this formulation, the phase field $\phi$ takes values from -1 (liquid) to 1 (vapor). The same equations for $\dot{\text{m}}$ and $\delta$ are used by Jafari \textit{et al.} to successfully simulate a growing vapor bubble in a confined tube \cite{Jafari2015}.

\renewcommand{\arraystretch}{1.3}
\begin{table*}
    \centering
    \begin{tabular}{|| c | c | c | c ||} 
    \hline
    \textbf{Parameter}  &  \textbf{Expression} &  \textbf{units} & \textbf{Description} \\ 
    \hline\hline
    $Mw$ & 0.018 & kg/mol & molecular weight of water \\  
    \hline
    $L$ & 42000/$Mw$ & J/kg & latent heat of vaporization \\ [0.5 ex] 
    \hline
    $T_{sat}$ & 373 & K & saturation temperature \\ 
    \hline
    $\rho_l$ & 1000 & kg/m$^3$ & density of water \\ 
    \hline
    $\rho_v$ & ($p_0\cdot Mw$)/(8.314$\cdot(T$+eps)) & kg/m$^3$ & density of water vapor\\ 
    \hline
    $R_T$ & 0.001 & m/s & tuning parameter \\ 
    \hline
    $hmax$ & 195  & $\mu$m & maximum grid size \\ 
    \hline
    $p_0$ & 101325  & Pa & atmospheric pressure \\ 
    \hline
    $\mu_l$ &  $2.82 \times 10^{-4}$ & Pa $\cdot$ s & dynamic viscosity of the fluid \\ 
    \hline
    $\mu_v$ &  $1.3 \times 10^{-4}$  & Pa $\cdot$ s & dynamic viscosity of the gas \\ 
    \hline
    $\gamma$ &  0.0588  & N/m & surface tension coefficient \\ 
    \hline
    $k_l$ & 0.63 & W/(m$\cdot$K) & thermal conductivity of water \\ 
    \hline
    $k_v$ & T$\times 8.32 \times 10^{-5} -7.46 \times 10^{-3}$  & W/(m$\cdot$K) & thermal conductivity of vapor\\ 
    \hline
    $k_T$ & $(k_l-k_v) \cdot V_{fl}+k_v$ & W/(m$\cdot$K) & thermal conductivity for two-phase flow\\ 
    \hline
    $Cp_l$ & 4200 & J/(kg$\cdot$K) & specific heat of liquid water \\ 
    \hline
    $Cp_v$ & 1840 & J/(kg$\cdot$K) & specific heat of water vapor \\ 
    \hline
    $Cp_f$ & $(Cp_l-Cp_v) \cdot V_l+Cp_v$ & J/(kg$\cdot$K) & specific heat for two-phase flow\\ 
    \hline
    
\end{tabular}
    \caption{Liquid and vapor simulation variables used in the COMSOL and lubrication models.}
    \label{Parameters}
\end{table*}

A weak expression, $W_k$, in the fluid domain helps produce smooth results. $W_k$ = test($\psi$) $\phi_s$+test($p$) $u_s$, where test() is a localized sampling function, $\phi_s = -\dot{\text{m}} \delta (V_{fv}/\rho_v + V_{fl}/\rho_l)$ is the source term of the phase field equation, $u_s = \dot{\text{m}} \delta (1/\rho_v - 1/\rho_l)$ is the source term in the continuity equation, and $V_{fv}$ is the volume fraction of the vapor. The phase field help variable is 
\begin{equation}
    \psi = -\vec{\nabla} \cdot \epsilon^2_{pf} \vec{\nabla} \phi + (\phi^2 -1) \phi +\frac{\epsilon^2_{pf}}{\lambda} \frac{\partial f}{\partial \phi},
\end{equation}
 where $\epsilon_{pf}$ is the interfacial thickness variable and $\lambda = 3 \epsilon_{pf} \sigma/\sqrt{8}$. The help variable describes the interfacial free energy between the two phases \cite{Patel2019, Jafari2015}. All other variables are defined in Table\ \ref{Parameters}.

For all simulations, we assume properties of the liquid and vapor phase associated with liquid and gaseous water. However, we used a larger value for the viscosity of the water vapor since it was necessary for numerical stability (as discussed in Sec.\ \ref{Results}). A typical initial phase field frame is shown in Fig.\ \ref{SimulationSize}a. For each simulation, initialization included defining the thermal properties of the hot solids, Table\ \ref{Table1}, and the starting isothermal temperature of all phases. For most cases, 650 K was used as the starting solid temperature. For some simulations, an increased solid temperature was required for large values of $H$ and low thermal conductivity materials. The initial water and vapor temperatures were set to 90$^{\circ}$C and 110$^{\circ}$C, respectively. 

\begin{table}[h]
    \centering
    \begin{tabular}{||l c| c | c | c ||} 
    \hline
    \textbf{Material} & &  $k_s \frac{W}{m \cdot K}$ &  $\rho \frac{kg}{m^3}$ & $C_p \frac{J}{kg \cdot K}$ \\ [0.5 ex] 
    \hline\hline
    \textbf{Copper} & Cu & 386 & 8940 & 385 \\ 
    \hline
    \textbf{Aluminum} & Al & 247 & 2710 & 897 \\
    \hline
    \textbf{Iron} & Fe & 60 & 7800 & 449 \\
    \hline
    \textbf{Titanium} & Ti & 7 & 4420 & 540 \\
    \hline
    \textbf{Glass} &   & 0.8 & 2500 & 792 \\
    \hline
\end{tabular}
    \caption{Thermal properties of the simulated hot object.}
    \label{Table1}
\end{table}

\begin{figure*}
    \centering
    \includegraphics[width= 6in]{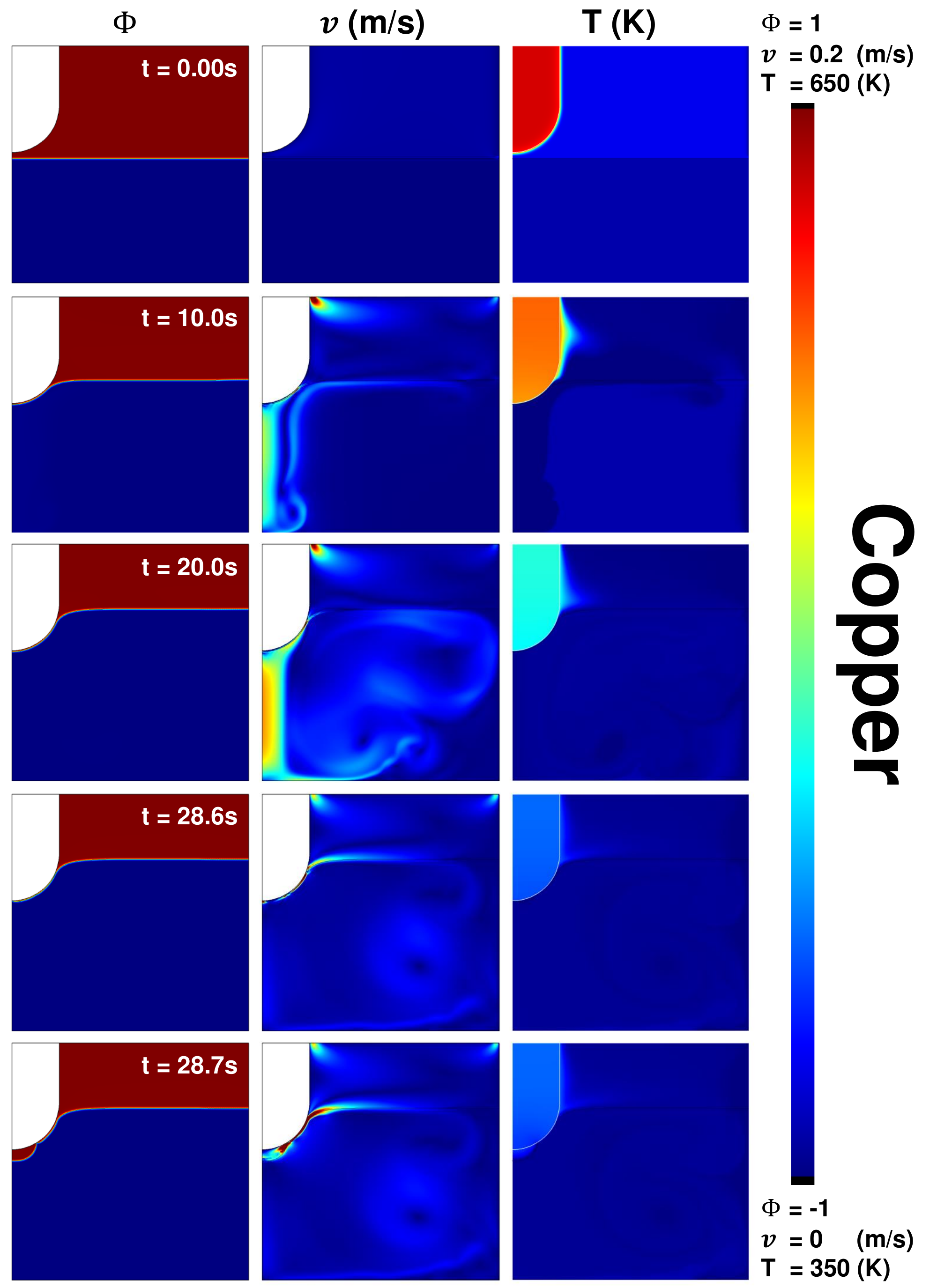}
    \caption[Copper Simulation Time Series]{A time series of an $R$ = 8 mm Copper object from top to bottom. From left to right is a phase field, velocity, and temperature field at the same time. The color map is $-1 \leq \phi \leq 1$, $0 \leq v \leq 0.2$ (m/s), and $350 \leq T \leq 650$ K for columns 1-3 respectively. The vapor layer fails between frames 28.6 and 28.7s.}
    \label{Copper}
\end{figure*}

\begin{figure*}
    \centering
    \includegraphics[width=6in]{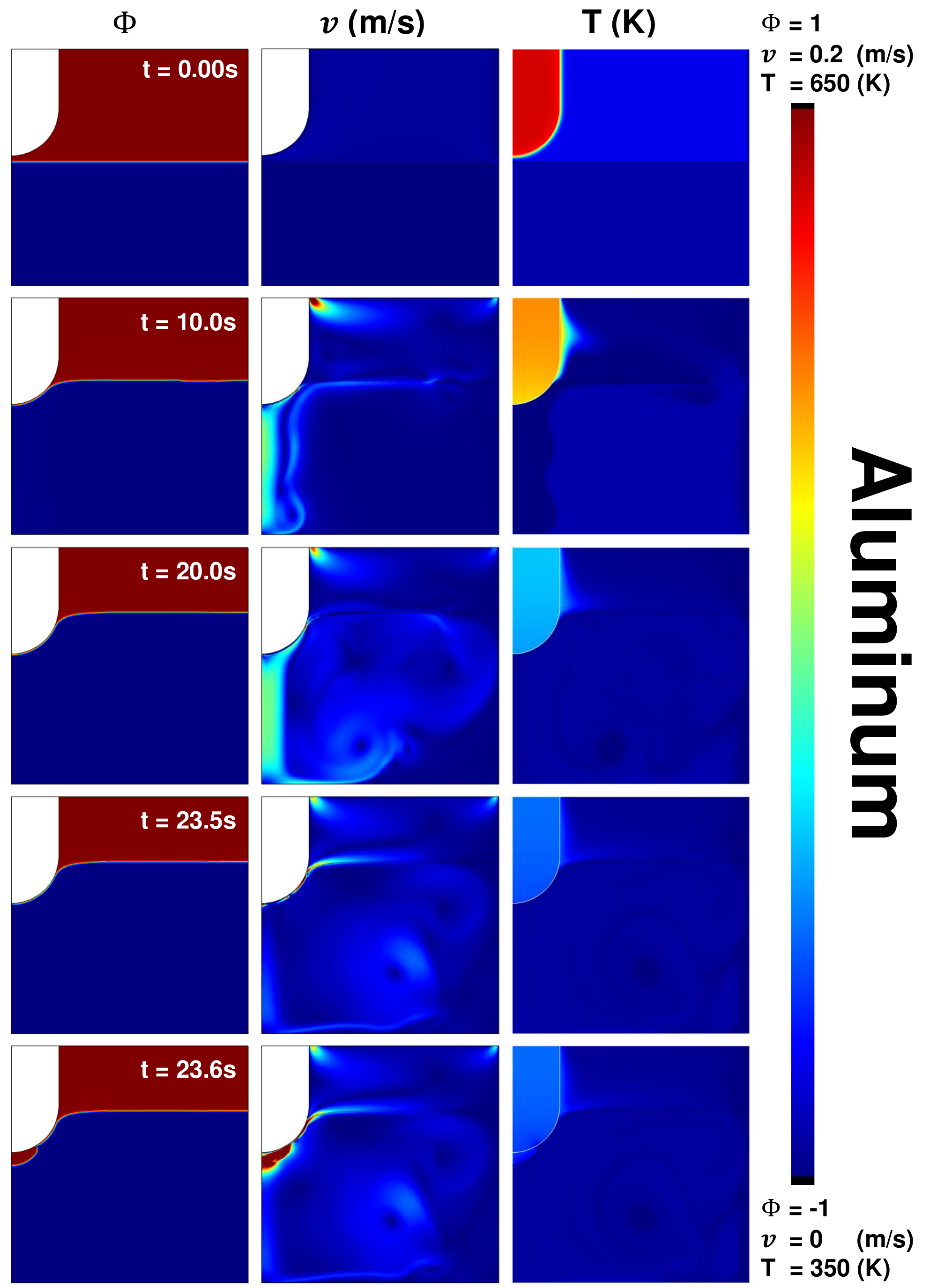}
    \caption[Aluminum Simulation Time Series]{A time series of an $R$ = 8 mm Aluminum object from top to bottom. From left to right is a phase field, velocity, and temperature field at the same time. The color map is $-1 \leq \phi \leq 1$, $0 \leq v \leq 0.2$ (m/s), and $350 \leq T \leq 650$ K for columns 1-3 respectively. The vapor layer fails between frames 23.5 and 23.6s.}
    \label{Aluminum}
\end{figure*}

\begin{figure*}
    \centering
    \includegraphics[width=6in]{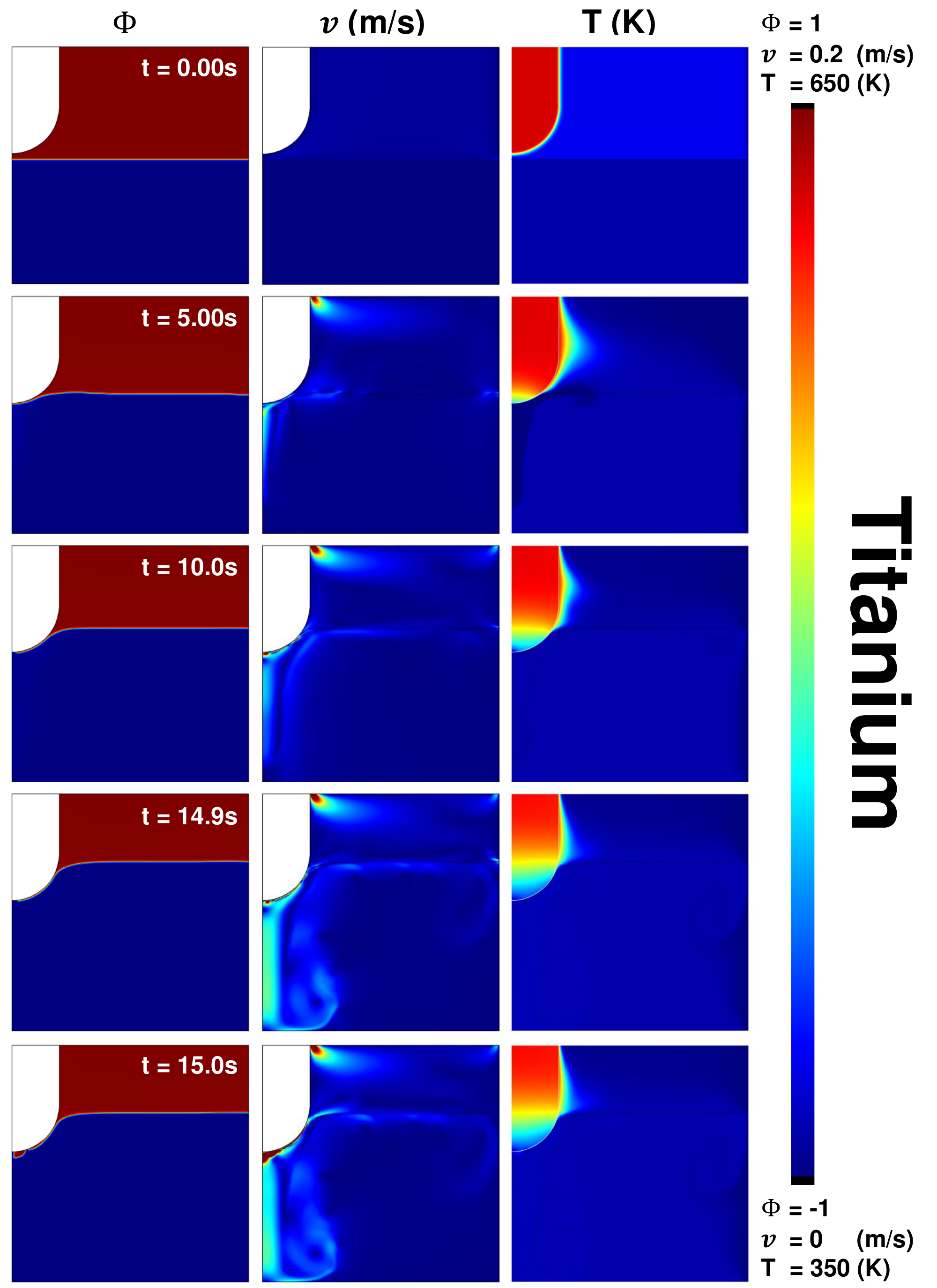}
    \caption[Titanium Simulation Time Series]{A time series of an $R$ = 8 mm Titanium object from top to bottom. From left to right is a phase field, velocity, and temperature field at the same time. The color map is $-1 \leq \phi \leq 1$, $0 \leq v \leq 0.2$ (m/s), and $350 \leq T \leq 650$ K for columns 1-3 respectively. The vapor layer fails between frames 14.9 and 15.0s.}
    \label{Titanium}
\end{figure*}

An atmospheric pressure outlet at $z$ = 40 mm allowed gas to escape, while an inlet at $z$ = 0 mm had a variable pressure, $P$, which controlled the equilibrium water level $H$. Initially, $P$ was defined and maintained for 1 s at $P = P_0 = \rho_l g z$, where $z$ = 21 mm. After 1 s of simulation time, the pressure was slowly increased according to $P = 5*t+P_0$, which raised the water height until $P = P_0 + \rho_l g (H $ + 1 mm), as shown in Fig.\ \ref{SimulationSize}b. As the water level rose a vapor layer spontaneously formed around the hot object due to evaporation. The pressure boundary at $z = 0$ mm conveniently minimized the effects of using the non-conservative mass equations by maintaining $H$ with as much water inflow as was necessary. Once the desired $H$ was achieved, the simulations continued unhindered until the vapor layer failed. Time series for typical simulations of copper, aluminum, and titanium solid objects with $R = 8$ mm and $H = 8$ mm, are shown in Figs.\ \ref{Copper}-\ref{Titanium}, respectively. These simulations will be discussed further in Sec.\ \ref{Results}.

\section{Lubrication Model and Comparison} \label{LubeSecs}

To help ensure the validity of the COMSOL Multiphysics simulations, we compared quiescent, near steady-state profiles at early times with a steady-state lubrication approximation. Nearly identical lubrication models have been used to describe the air or evaporated vapor flow under a levitated drop \cite{lister2008shape,Snoeijer2009,duchemin2005static,Chakraborty2022, Sobac2017, Sobac2014}. In particular, predictions from Snoeijer \textit{et al.} \cite{Snoeijer2009} show excellent agreement with experiments of Leidenfrost drops \cite{Burton,Biance}. The lubrication approximation balances the leading order pressure gradient terms driving the tangential flow in the vapor layer with gradients in viscous stress, and ensures conservation of mass in the flow. For the lubrication model presented here, both the liquid-vapor interface and the solid material are considered motionless. This is valid if the viscosity of the liquid is considerably larger than the vapor. Although in our COMSOL simulation the liquid is only twice as viscous as the vapor (Table \ref{Parameters}), this simplification allows one to ignore flow in the liquid phase and serves as a leading-order description of the physics. 

\begin{figure}
    \centering
    \includegraphics[width=\columnwidth]{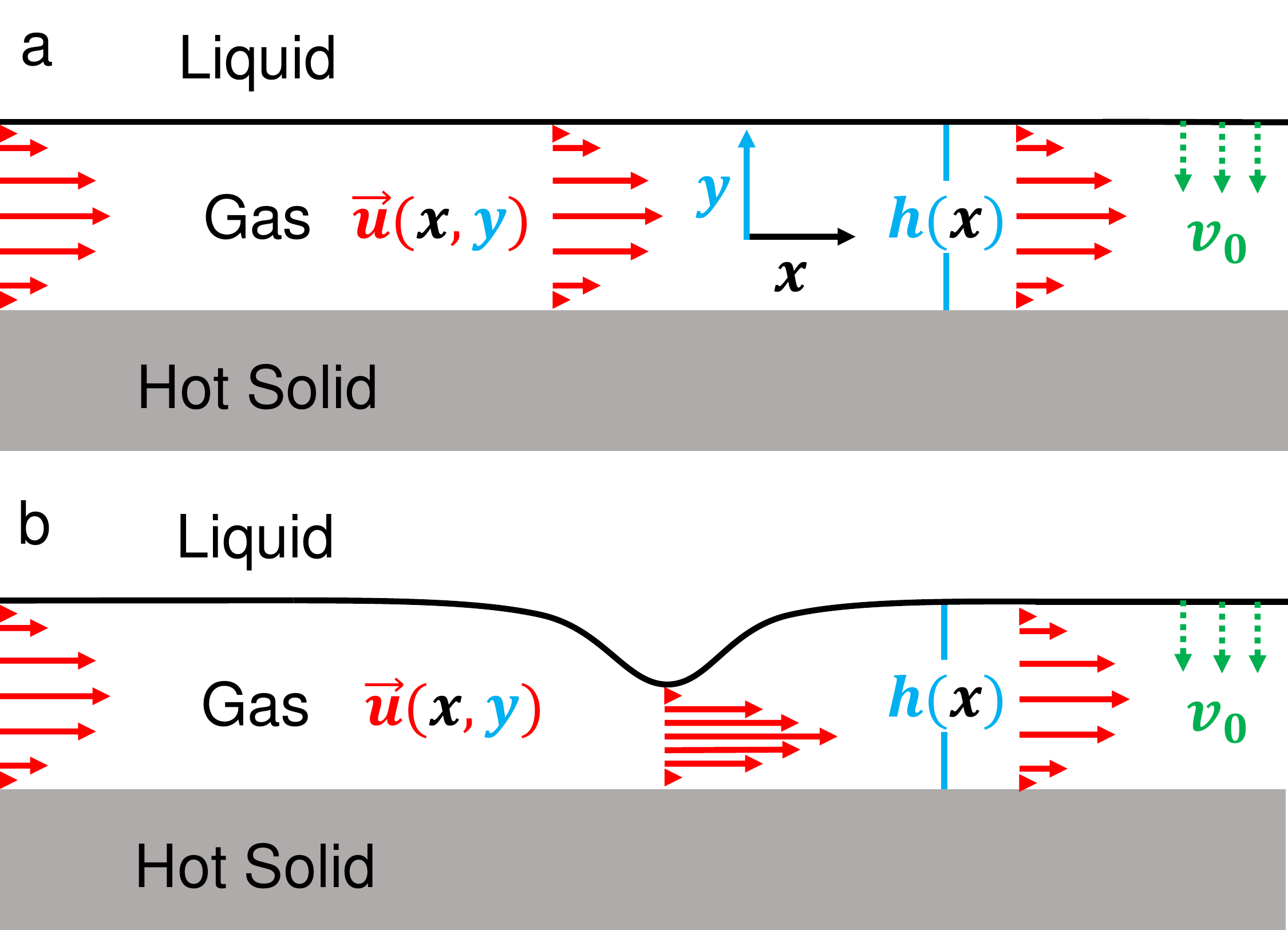}
    \caption[Lubricating Flow]{(a) Schematic of a time-independent 2D lubrication approximation of vapor flow between a hot solid and a fluid. Here $h$ is the thickness of the vapor layer along $x$, and $\vec{\bf u}(x,y)$ is the velocity field. The evaporation feeds the gas layer at a velocity $v_0$. In this system $dh/dx << 1$ and a parabolic flow profile allows the velocity to be integrated over $y$, simplifying $\vec{\bf u}(x,y)$ to $\bar{u}(x)$. (b) Prior to vapor layer failure, a perturbation forms and grows so that $dh/dx\approx 1$, meaning that lubrication theory isn't strictly valid. %where now you must consider the flow between the wavepeak and the solid. The wavelength, which becomes the characteristic length scale $L$ for fluid between the peal and the solid, is comparable to $h$ so $h/L << 1$ is not true.
    }
    \label{LubricationFlow}
\end{figure}

In general, the lubrication approximation makes two major assumptions to simplify the governing equations. First, $h/L << 1$, where $h$ is the local gap thickness and $L$ is the total vapor path length. This is valid in steady-state Leidenfrost drops where $10 \leq h \leq 100$ $\mu$m and $L \leq2$ cm. Second, the vapor has a parabolic velocity profile within the gap such that $u_x(x,y,t) = 6\bar{u}y(h-y)/h^2$, and $\bar{u}=\bar{u}(x,t)$ is the mean velocity in the tangential direction \cite{lautrup2004physics}. A schematic of this geometry is shown in Fig.\ \ref{LubricationFlow}a. A further simplification is typically made by assuming steady-state, so that $\bar{u}=\bar{u}(x)$. Although this results in a system of 2 ODEs for $\bar{u}(x)$ and $h(x)$, during failure of the vapor layer these assumptions break down. The wavelength and amplitude of the instability, which causes vapor layer failure, grow in time so that $dh/dx\approx1$, Fig.\ \ref{LubricationFlow}b. The wavelength becomes the important longitudinal length scale, $L$, thus breaking the assumption that $h/L << 1$. Subsequently, we only can compare our simulations to the lubrication approximation at steady-state times well before failure.

\begin{figure}
    \centering
    \includegraphics[width=\columnwidth]{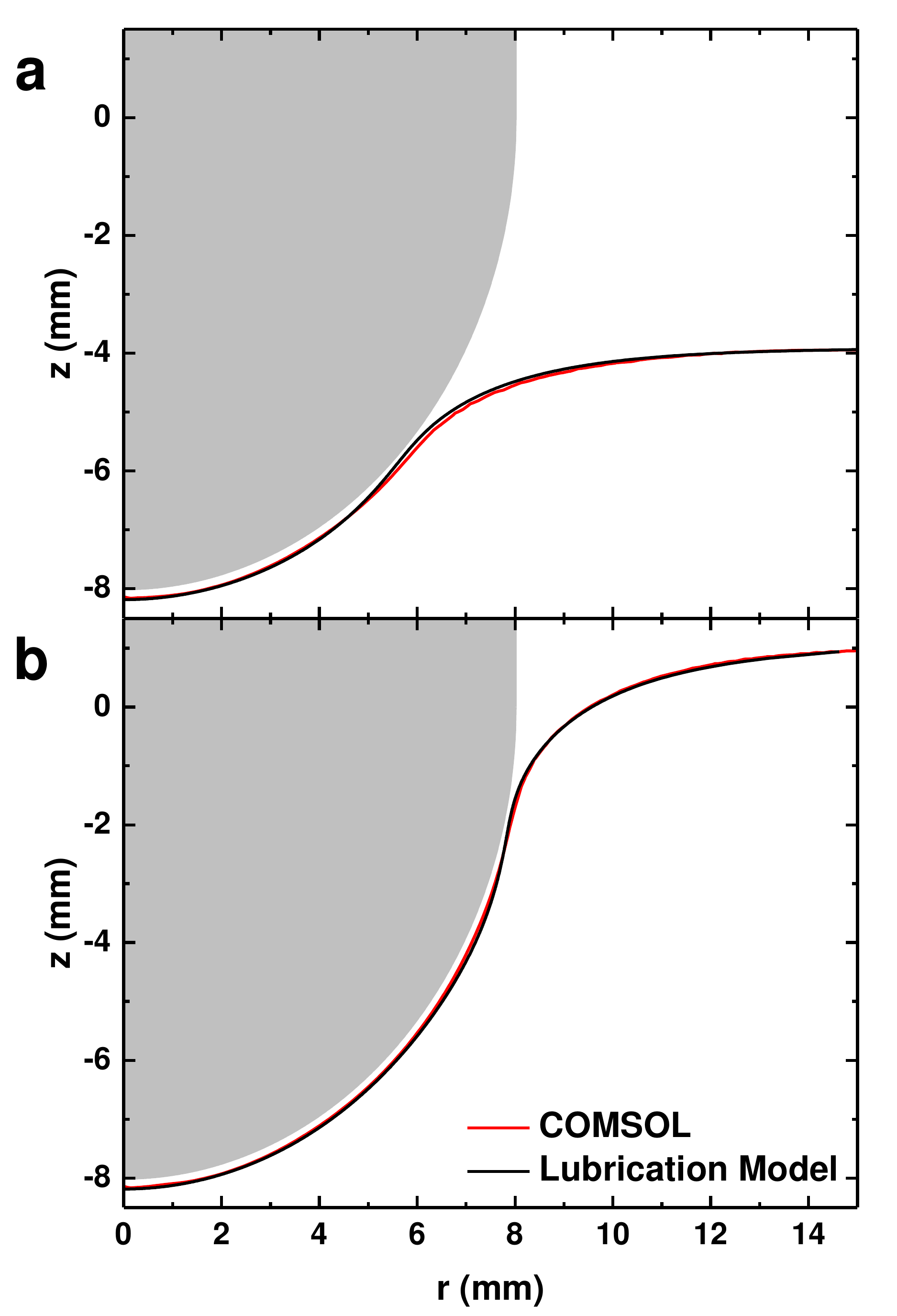}
    \caption[COMSOL \& Lubrication Interface Comparison]{Comparison between the COMSOL rendered interface for $\phi = 0$ at steady-state, red, and the lubrication modeled interface, black. Both models use the parameters in Table \ref{Parameters}. For (a), $H$ = 5 mm, $h(\pi)$ = 185.3 $\mu$m, $h''(\pi)$ = 402.5 $\mu$m, and $v_\text{0}$ = 9.31 mm/s. For (b), $H$ = 10 mm, $h(\pi)$ = 185.2 $\mu$m, $h''(\pi)$ = 33.1 $\mu$m, and $v_\text{0}$ = 9.97 mm/s.}
    \label{ContourCompare}
\end{figure}

\begin{figure}
    \centering
    \includegraphics[width=\columnwidth]{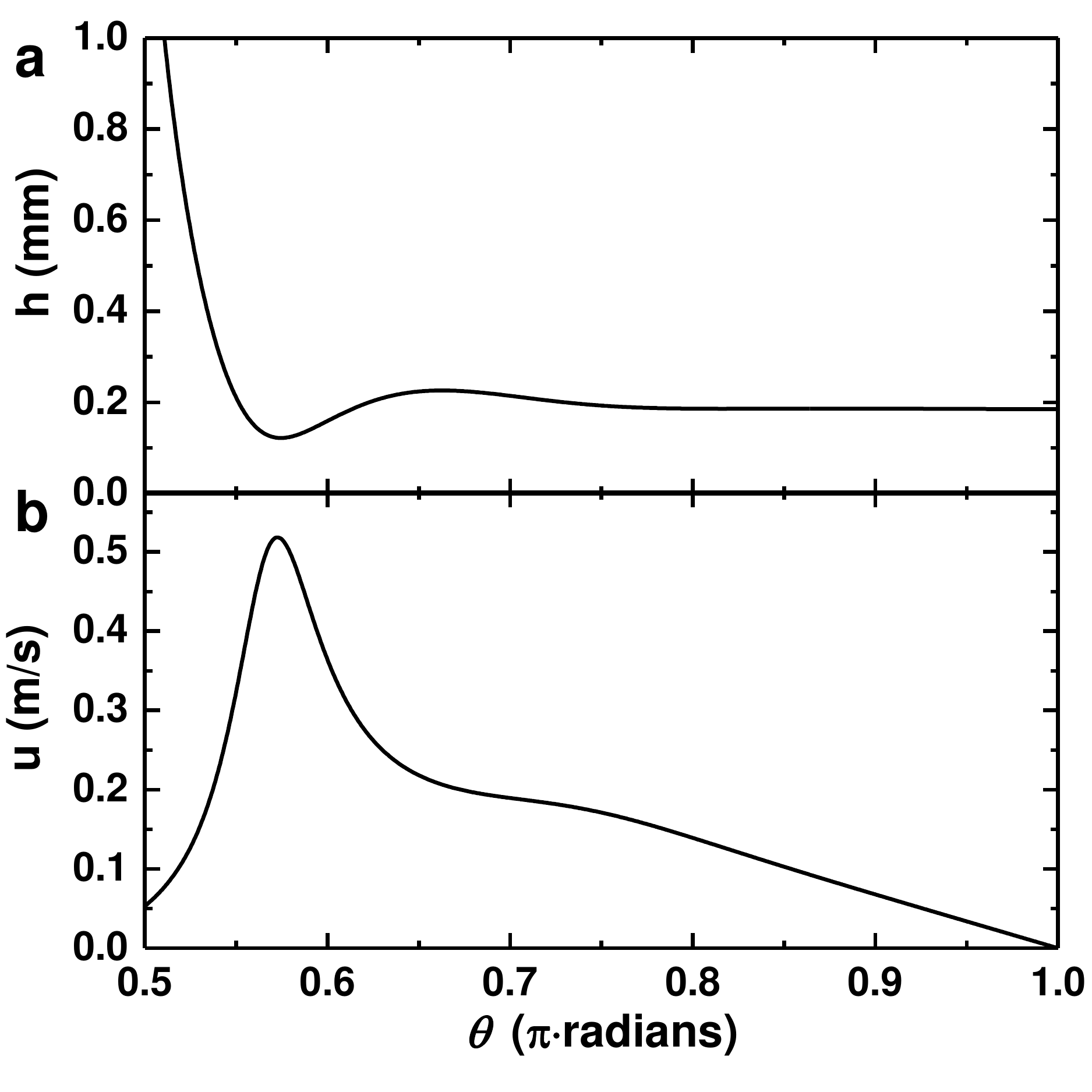}
    \caption{Solution of Eqs. \ref{masscon} and \ref{pcon} for (a) $h$ and (b) $\bar{u}$ using the boundary conditions $h(\pi)$ = 185.2 $\mu$m, $h''(\pi)$ = 33.1 $\mu$m, and $v_\text{0}$ = 9.97 mm/s. These solutions correspond to Fig.\ \ref{ContourCompare}b where $\theta = \pi/2$ is the vapor flow outlet and $\theta = \pi$ is the tip of the hot object.}
    \label{Supp7}
\end{figure}

In our lubrication model, we ignore inertia. Estimates for the flow beneath centimeter-scale droplets suggest that Re $\lesssim$ 0.3 \cite{Ma}, so this assumption is reasonable. Since we are not explicitly considering evaporation, we assume that there is a constant flow of gas that feeds the vapor layer at velocity $v_\text{0}$. We only consider the flow around the lower part of a hemisphere immersed in the liquid, and use axisymmetric spherical coordinates so that the tip of the hemisphere (with radius $R$) corresponds to $\theta=\pi$ (Fig.\ \ref{ExperimentDiagram}a). Both the film thickness, $h(\theta)$, and the depth-averaged velocity, $\bar{u}(\theta)$, vary with the angular coordinate. We will assume that $h\ll R$, and that $h'\ll R$, where prime denotes differentiation with respect to $\theta$. This means that the film is thin and that its thickness varies slowly along the edge of the hemisphere, an assumption that must break down at failure.

With these assumptions, there are two differential equations that describe the steady state profiles of $h$ and $\bar{u}(\theta)$:
\begin{eqnarray}
&h\bar{u}\sin\theta=-Rv_\text{0}(1+\cos\theta),\label{masscon}\\
&12 \eta\dfrac{\bar{u}}{h^2}=\dfrac{1}{R}\dfrac{d}{d\theta}\Big[\rho_\text{l}g(R+h)\cos\theta+\gamma\kappa\Big].\label{pcon}
\end{eqnarray}
Equation\ \ref{masscon} comes from mass conservation of the gas in the vapor layer, assuming that $\bar{u}(\pi)=0$ from symmetry considerations. 
%Mass continuity in the vapor layer at $\theta$ results from the integrated amount of gas entering the vapor layer, starting from $\theta=\pi$. 
Equation\ \ref{pcon} balances the pressure gradient necessary to drive the flow $\bar{u}$ in the lubricating vapor layer with the pressure gradient determined by hydrostatic (gravitational) forces and variations in the Laplace pressure. The variable $\kappa=\kappa(\theta)$ is the total curvature of the interface, and can be calculated by taking the divergence of the normal vector to the liquid-vapor interface:
\begin{eqnarray}
&\kappa=\vec{\nabla}\cdot\hat{\bf n},\\
&\hat{\bf n}=\left(-\hat{\bf r}+\dfrac{h'}{R+h}\hat{\boldsymbol \theta}\right)\left(1+\dfrac{h'^2}{(R+h)^2}\right)^{-1/2},
\end{eqnarray}
where the spherical coordinate $r$ is evaluated at $R+h$ after differentiation. 
Here we have kept the full curvature with no approximations in order to correctly model the shape of the interface all the way to the boundary.

Equation\ \ref{masscon} can be solved directly for $\bar{u}$ and inserted into Eq.\ \ref{pcon}, resulting in a third-order differential equation for $h(\theta)$. For a given gas input velocity ($v_\text{0}$), we chose initial values of $h(\pi)$ and $h''(\pi)$ since $h'(\pi)$ = 0 by axisymmetry. We then solved the resulting equation for $h(\theta)$ until we matched the height of the liquid level at the outer boundary ($r\gg R$), and required that $h'=(R+h)\cot\theta$, so that the normal to the interface pointed in the $z$-direction. Importantly, we emphasize that these are steady solutions, so there are no dynamics. This means that for a given $v_\text{0}$ and set of boundary conditions, a solution may not exist. In fact, the structure of the branches where solutions exist is quite complicated, as may be expected from similar problems considering the lubricating flow under droplets \cite{lister2008shape,duchemin2005static}. In reality, a vapor layer may exist for all thicknesses and boundary conditions, but will fluctuate in time.
%In reality, a vapor layer may exist for a given thickness and boundary conditions in the experiments, but it will fluctuate in time. 
Periods of quiescent behavior in the experiments \cite{Harvey2021} and COMSOL simulations may correspond to regimes where steady solutions of the lubricating vapor layer exist. 
% There are a few things to note about the solutions. First, the solution is technically not valid when $h\sim R$. For boundaries which are not too far from the solid edge at $R$, this is acceptable \cite{lister2008shape}, but for far-field boundaries the solution would need to be matched to an outer solution for the liquid-vapor interface. 

Using the properties for water and vapor given in Table \ref{Parameters}, we show comparisons between the lubrication model and COMSOL simulation at early times in Fig.\ \ref{ContourCompare}. The lubrication model parameters $h(\pi)$, $h''(\pi)$, and $v_\text{0}$ were adjusted to best match the COMSOL profiles where the phase field $\phi=0$ (50\% vapor, 50\% liquid). The agreement is excellent for different values of the external water level $H$. However, the lubrication model tends to have a minimum in $h$ that is smaller than in the COMSOL simulation, meaning that the interface comes closer to the solid boundary. This may be due to the variation in evaporation rate in the COMSOL model, which is determined by thermodynamics, whereas the lubrication model assumes a constant value of $v_\text{0}$ along the entire interface. Figure \ref{Supp7} shows the solutions for $h(\theta)$ and $\bar{u}(\theta)$ that correspond to the profile shown in Fig.\ \ref{ContourCompare}b. The minimum in $h$ aligns with the maximum in the velocity $\bar{u}$. The near linear increase in the velocity from $\theta=\pi$ is a result of the increasing mass flux into the vapor layer (Eq.\ \ref{masscon}). The peak in $|\bar{u}|\approx$ 0.5 m/s agrees very well with the velocity from the full COMSOL simulation, discussed in more detail in Sec.\ \ref{Results}.

%The problem is richer than what is presented here, yet a detailed examination of the solution branches and their stability is left for future work.

\section{Failure of the Vapor Layer} \label{Results}

\begin{figure*}
    \centering
    \includegraphics[width= 6.5 in]{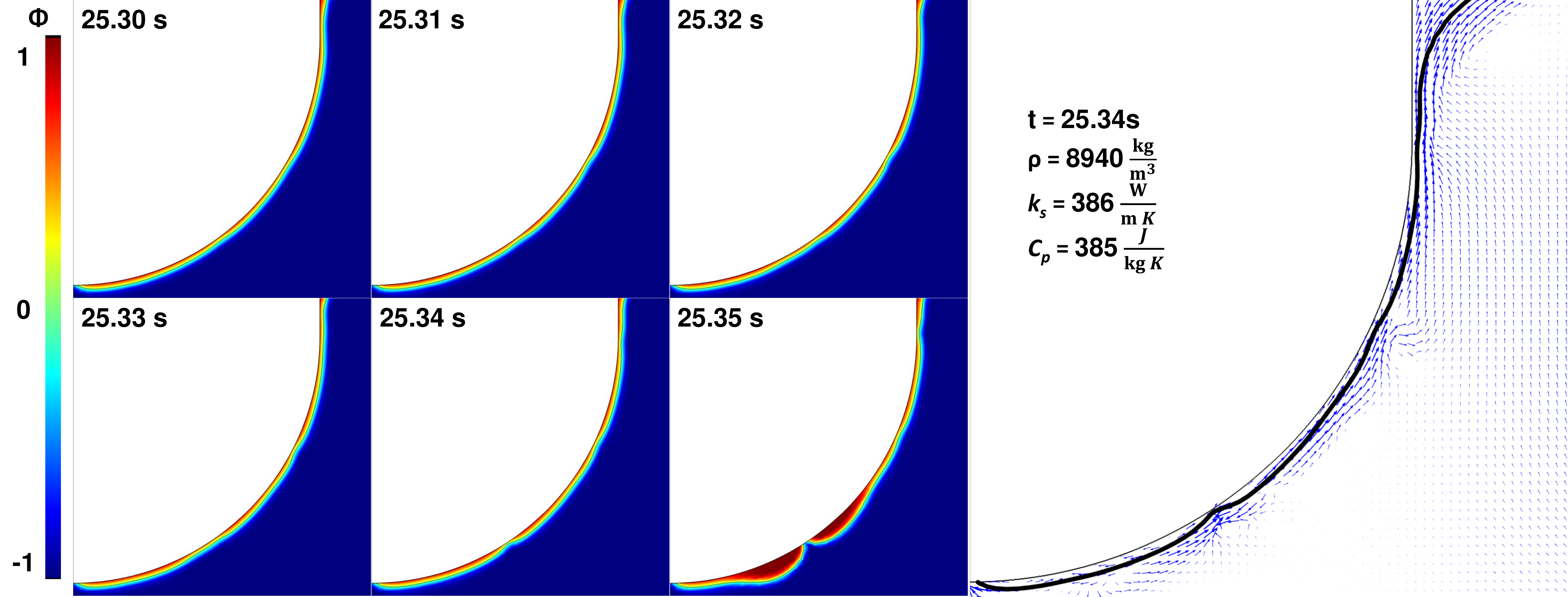}
    \caption[Simulation Wave Mechanism]{The left 6 frames show a phase field time series leading to the failure of the vapor layer near a solid with the thermal properties of copper. Surface waves are clearly visible and grow in time until one peak contacts the solid. On the right is a velocity field 10 ms before vapor layer failure.}
    \label{SimulationTimeSeries}
\end{figure*}

\begin{figure*}
    \centering
    \includegraphics[width= 5.2 in]{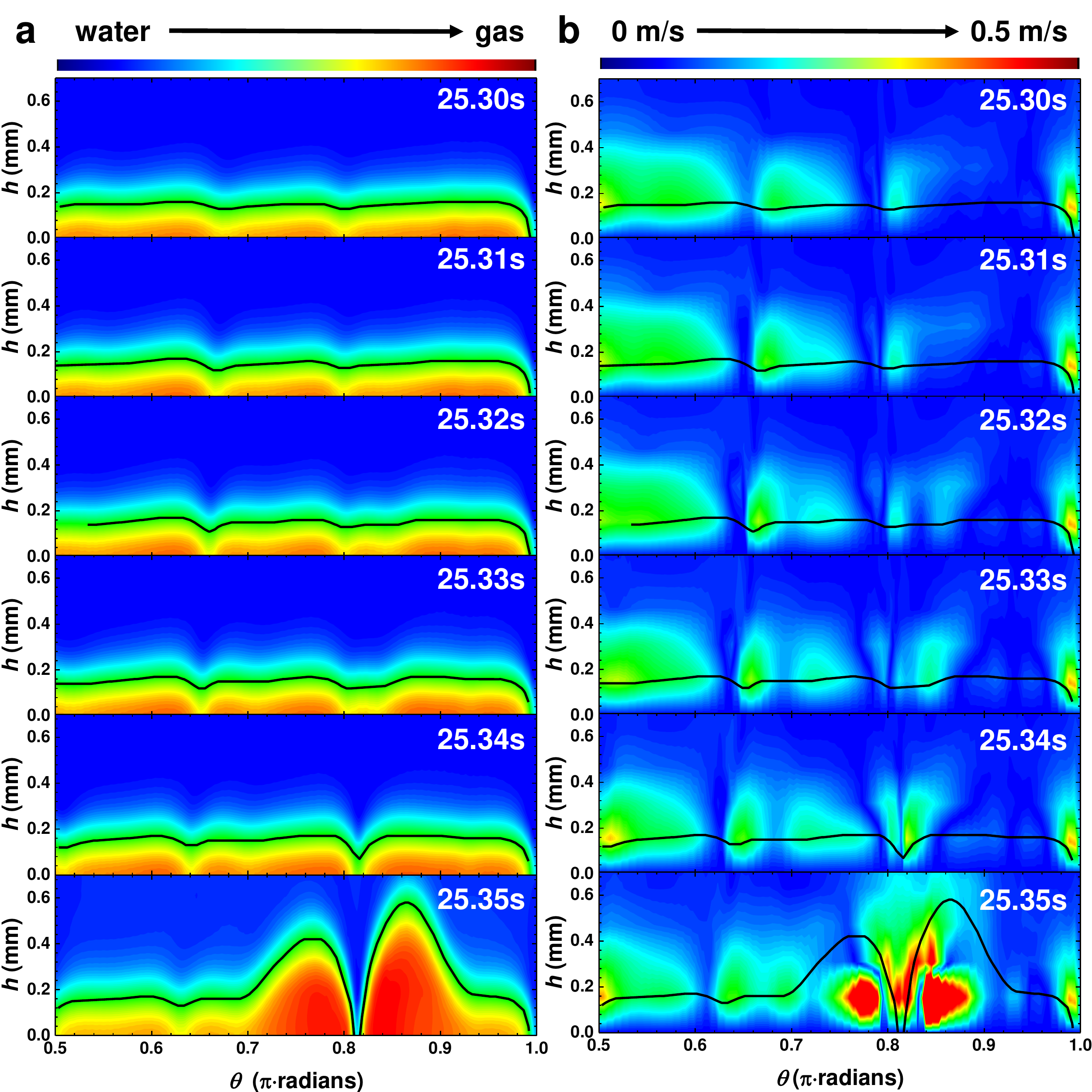}
    \caption{Time series of vapor layer failure (unwrapped in the angular coordinate) of (a) the phase field and (b) the velocity magnitude. The tip of the hot object is located at $\theta$ = $\pi$ and gas escapes above $\theta$ = $\pi$/2. Here we use the thickness $h$ as a coordinate to represent the distance away from the hot object. The black line is the $\phi$ = 0 contour of the gas-fluid interface. As time progresses, multiple perturbations evolve, yet only one leads to failure at a single point.}
    \label{Unwrapped1}
\end{figure*}

The primary purpose of the COMSOL simulations was to provide a quantitative understanding of the failure of Leidenfrost vapor layers. As such, we varied the geometry of the vapor layer and the thermal properties of the hot solid to best match our previous experimental results \cite{Harvey2021}. In experiments, we found the minimum Leidenfrost Temperature, $T_-$ = 140$^\circ$C, was independent of solid thermal properties, geometry, and salt concentration of the liquid. However, the physical geometry of the metallic hot solid was limited by experimental constraints. Here, we show how our simulations can replicate experimental results, explore a larger variation of system parameters, and provide insight into the instability that drives vapor layer failure.

%\begin{figure}
 %   \centering
  %  \includegraphics[width = 3.25 in]{DefineSurfaceArea.pdf}
  %  \caption[Surface area of the vapor layer]{The effective surface area is defined as the surface integral of the hemi-sphere to the inflection point of the vapor layer interface defined as a. }
  %  \label{SurfaceArea}
%\end{figure}

A typical simulation evolved as described in Sec.\ \ref{ComsolDetails} until vapor layer failure (explosive collapse). As the water level rose, evaporation produced a vapor layer of order 100-200 $\mu$m around the hot solid object. During subsequent cooling, for solids such as aluminum or copper with high thermal conductivity, the solid was nearly isothermal during the simulation (Figs.\ \ref{Copper} \& \ref{Aluminum}). However, there was significant local cooling near the tip of the solid in materials with low thermal conductivity, such as titanium (Fig.\ \ref{Titanium}). In all simulations, local cooling of the liquid due to evaporation, determined by $Q_s$, was initially strong enough to create a downward convective flow, as illustrated in Figs.\ \ref{Copper}-\ref{Titanium}. This downward flow eventually subsided as the solid cooled, but persisted until failure for titanium. As a consequence, the flow in the liquid adjacent to the vapor layer could reverse direction during the simulation.

Upon further cooling of the solid, the vapor layer thinned and developed time-dependent variations in thickness; a dynamic instability leading to failure. In experiments, this instability manifested as visible surface waves, Fig.\ \ref{ExperimentDiagram}c, which proceeded the explosive collapse of the vapor layer. In simulations, the instability appeared as the same interfacial wave, sometimes many waves, which grew and moved along the surface until fluid-solid contact (last frames of Figs.\ \ref{Copper}-\ref{Titanium}). This is most clearly illustrated in Fig.\ \ref{SimulationTimeSeries} (Movie S3), which shows a times series of vapor layer failure for a copper object. The location of the failure point was somewhat stochastic and typically did not occur at the minimum in $h$ (\textit{e.g.}, Fig.\ \ref{Supp7}a). 

Figure \ref{Unwrapped1} shows an unwrapped angular plot of both the phase field and velocity magnitude from the same simulation. An obvious feature seen here is the appearance of local liquid contact at the tip, $\theta=\pi$. This occurred in nearly all simulations, but most prominently for solids with lower thermal conductivity and small values of $R$. This apparent contact did not produce explosive boiling, and the vapor layer failure seemed to occur at smaller values of $\theta$ away from the tip. Additionally, this apparent contact was not observed in experiments \cite{Harvey2021}, and is likely a consequence of the finite thickness of our phase field or the enforced axisymmetric geometry. Nevertheless, at the failure point away from the tip, liquid-solid contact occurred in 10s of milliseconds, which agrees well with the rapid timescales for collapse measured in experiments \cite{Harvey2021}. At the contact point, the temperature of the solid, $T_s$, was still well above the boiling point of water, so rapid evaporation ensued at the moment of contact. Subsequently, the COMSOL simulation often crashed due to the explosive nature of the event and the inability to capture such rapid dynamics. However, we can analyze the results of the simulation leading to failure and compare directly to experiments.

\begin{figure}
    \centering
    \includegraphics[width=\columnwidth]{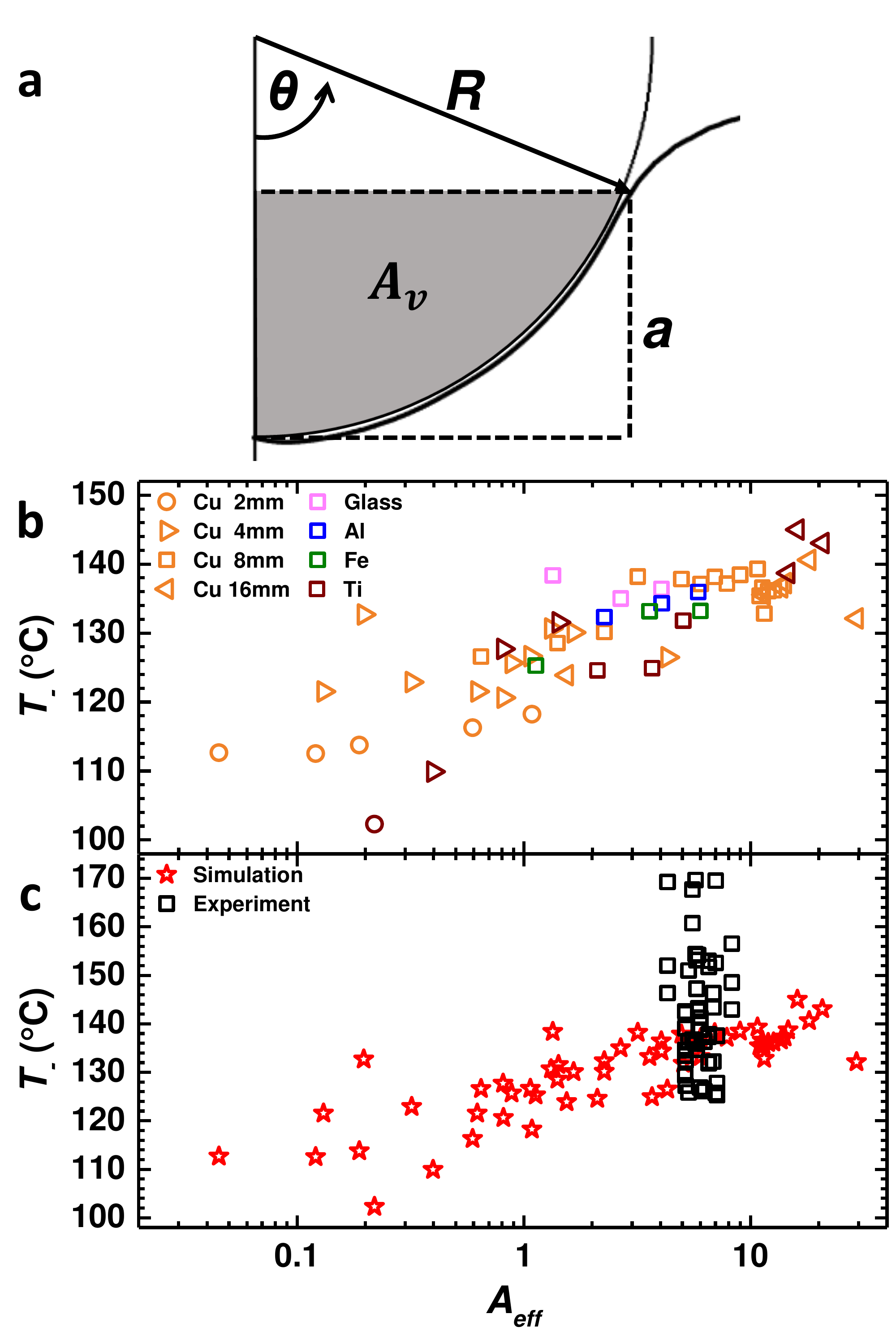}
    \caption[$T_-$ vs $A_eff$ of Simulations and Experiments]{
   (a) $A_{v}$ is the surface area of the shaded region depicted in the picture. Equation \ref{SurfaceEq} defines the area from the tip of the hot object to the inflection point of the fluid interface at height $a$. (b) Local failure temperature, $T_-$, for simulations using various material properties, denoted by color, and object radii $R=$ 2-16 mm, denoted by symbols, versus the normalized surface area of the vapor layer. $A_{eff}=A_{v}/A_{lc}$, as defined in the text. (c) Comparison of $T_-$ from all simulations with data from experiments \cite{Harvey2021}, where simulations act as a lower bound for experimental results.}
    \label{SimulationFailureTemp}
\end{figure}

In our experimental work, we could precisely determine the failure of the vapor layer in time as a discontinuity in the thickness measurement \cite{Harvey2021}. That discontinuity correlated to the wetting of the surface as liquid came into contact with the hot solid. To directly compare simulations and experiments we define vapor layer failure as the initial contact point between the fluid and solid. Therefore, $T_-$ was measured locally in the solid at the contact point one time-step before collapse. Furthermore, to compare among simulations and experiments with different values of the tip radius $R$ and water height, $H$, we computed the effective surface area of the vapor layers, $A_{eff} = A_v/A_{lc}$. Here $A_v$ is the surface area of the vapor layer found by integrating over a semi-sphere \cite{Boas2005}: 
\begin{equation}
A_{v}=\int_0^{2 \pi} \int_0^{\cos^{-1}{\frac{R-a}{R}}} R^2\sin{\theta}d\theta d\Phi=2 \pi a R.
\label{SurfaceEq}
\end{equation}
Here $a$ is the height above the tip measured at the inflection point of the $\phi = 0$ contour (Fig.\ \ref{SimulationFailureTemp}a). We divide by $A_{lc} = 2 \pi l_c^2$ to normalize the area, where $l_c = \sqrt{\gamma/\rho_l g}=2.5$ mm is the capillary length of boiling water. Using two parameters, the radius of the hot solid, 2 $\leq R \leq 16$ mm, and the water height, $0.5 \leq H \leq 17$ mm, an effective area range of $0.02 < A_{eff} < 12$ was achieved in simulations. In Fig.\ \ref{SimulationFailureTemp}b we plot $T_-$ as a function of $A_{eff}$ for hot objects of various radii and thermal properties. The thermal properties correlated to real material values described in Table \ref{Table1}.

\begin{figure*}
    \centering
    \includegraphics[width= 5.5 in]{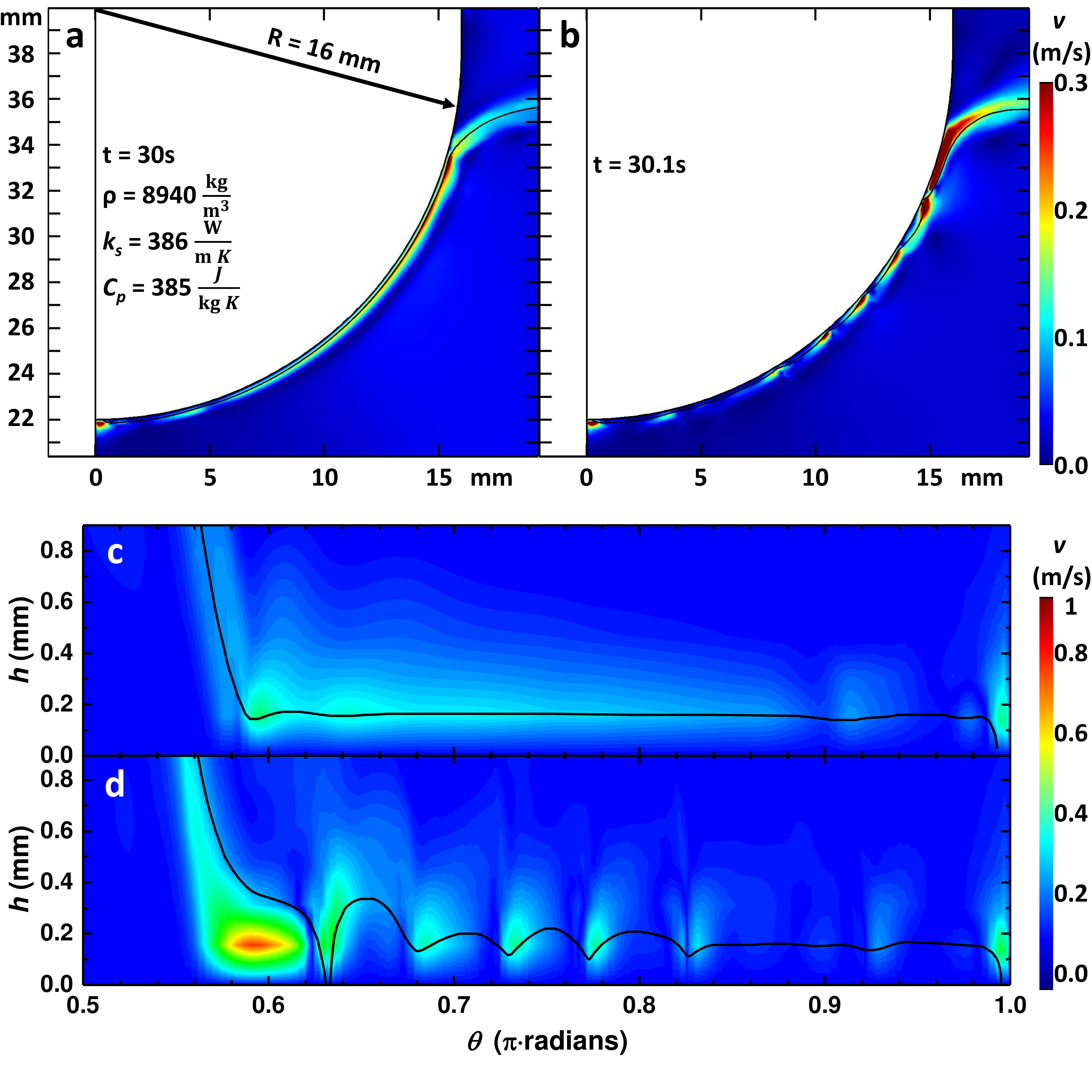}
    \caption[COMSOL Velocity Field]{Color map of the velocity magnitude of the two-phase fluid at 30 s (a) \& (c) and 30.1 s (b) \& (d) with the $\phi=$ 0 contour in black. The radius of the object was $R$ = 16 mm. (a) \& (b) show the geometry in cylindrical coordinates, while (c) \& (d) uses unwrapped angular coordinates. The angle $\theta$ is defined in Fig.\ \ref{ExperimentDiagram}a. Here we use the thickness $h$ as a coordinate to represent the distance away from the hot object. Interface perturbations with $\lambda \approx 2.45$ mm grow between the two frames and failure occurs at the apex of the wave furthest to the right in panel (b) (left in panel (d)).}
    \label{MultiplePerturbations}
\end{figure*}

Just as in experiments (Fig.\ \ref{ExperimentDiagram}b), we found that $T_-$ was nearly independent of the thermal properties of the solid. Generally, $T_-$ decreased as $A_{eff}$ decreased and even approached the boiling temperature \cite{NASA,Harvey2021,Zaitsev}, yet $T_-$ appeared to plateau at larger values of $A_{eff}$. The plateau region begins at $A_{eff}\approx 1$, so that the characteristic size of the vapor layer is approximately the capillary length, $l_c$. This is consistent with experiments with liquid drops where the effective area could be easily varied by more than 1 decade \cite{Harvey2021}. In the experiments using an hot object immersed in a bath, as simulated here, the effective area was limited to a much smaller range, as shown in Fig.\ \ref{ExperimentDiagram}b. In this range of $A_{eff}$, the simulations act as a lower bound for experimental values of $T_-$. This should be expected since there were uncontrollable factors in the experiments such as microscopic debris in the water and finite surface roughness of the hot solid. Both of these effects could lead to premature failure at higher temperatures. Overall, it is remarkable that the simulations agree quantitatively with the experiments despite their simplifications, such as enforcing an axisymmetric geometry, and necessitating a higher vapor viscosity for numerical stability. This implies that the nature of the failure mechanism depends on other fluid properties of the vapor.

To better understand the evolution of failure, we investigated larger solid objects ($R=16$ mm) where multiple perturbations could grow, move, and evolve prior to failure. Figure \ref{MultiplePerturbations}a-b shows two final frames from the simulation. A relatively smooth vapor layer quickly developed a series of thickness and velocity perturbations resembling a wave train. The spacing between the perturbations decreased as $\theta$ decreased towards the vapor exit. Since the average vapor velocity also increases near the exit (\textit{i.e.}, Fig.\ \ref{Supp7}b), we surmise that the local growth of each perturbation depends on the local vapor velocity. To visualize this more clearly, we plot the unwrapped phase contour and velocity magnitude in Fig.\ \ref{MultiplePerturbations}c-d. Prior to the growth of the perturbations, the thickness profile agrees well with the prediction from the lubrication model, Fig.\ \ref{Supp7}a. Just before failure, there is a clear characteristic wavelength ($\lambda$) for the series of perturbations that develop. In units of length, this wavelength corresponds to $\lambda\approx 2.45$ mm. The appearance of a definitive wavelength suggests a hydrodynamic instability is initiated that leads to liquid-solid contact.

How does this hydrodynamic instability determine the failure temperature? In the vapor layer flow, there are two major forces that could naturally lead to an instability. First, the pressure in the vapor layer acts against buoyancy, so that a Rayleigh-Taylor (RT) type of instability could lead to liquid touching the surface. In fact, the RT instability is known to determine the maximum size of Leidenfrost drops \cite{Biance,Burton,Snoeijer2009}. Second, the rapid flow in the vapor layer applies a stress to the liquid-vapor interface and can lead to a Kelvin-Helmholtz type of instability. However, the simplest example of both of these instabilities lead to critical wavelengths of order 1-2 cm \cite{lautrup2004physics}, not millimeters, as we observe in our simulations and experiments \cite{Harvey2021}. In the Leidenfrost vapor layer, there are additional complexities that need to be considered, such as the presence of a solid surface, viscous and inertial forces, evaporation, and a non-planar geometry. 
%We must look elsewhere for the instability mechanism and consider all possible forces in our system that could be contributing to it.

Consider the simple picture of a wave at the gas-fluid interface depicted in Fig.\ \ref{LubricationFlow}b and the forces acting on it. Surface tension and evaporation both act as restoring forces to suppress the growth of liquid-vapor interface perturbations. For example, the rate of evaporation should increases as the liquid approaches the solid, producing a higher local pressure. 
%The surface tension wants to minimize the surface area, and thus restores any local perturbations that form. 
As mentioned, buoyant forces (gravity) can instigate the growth of perturbations (RT instability), but only for centimeter-scale wavelengths. The net effect of viscous lubrication pressure depends on the geometry of the perturbation. For a symmetric perturbation, a high and low pressure develop that are equal in magnitude on each side of the perturbation, resulting in no net lift. Thus, for an infinitesimal sinusoidal perturbation to a flat vapor interface, viscous forces alone may not explain this instability. 

We hypothesize that inertial forces in the vapor layer can lead to a reduction in pressure sufficient to draw the liquid-vapor interface close to the solid, initiating contact and failure. Indeed, inertia has been shown to be necessary in the formation of finite-time singularities in similar two-dimensional flows \cite{Dupont1993,Burton2007}. An overly simplistic balance of Laplace and Bernoulli pressure can illustrate this idea. Referring to Fig.\ \ref{LubricationFlow} again, consider a sinusoidal perturbation of amplitude $\Delta h$ and wavelength $\lambda$ to the flat liquid-vapor interface,
\begin{equation}
    h=h_0+\Delta h \sin(2\pi x/\lambda),
\end{equation}
where $h_0$ is the mean vapor layer thickness. Conservation of mass in the vapor layer leads to an increase in the velocity, $\Delta u$, where the perturbation is closest to the surface,
\begin{equation}
    h_0 u_0=(h_0-\Delta h)(u_0+\Delta u),
\end{equation}
where $u_0$ is the mean velocity in the vapor layer. The reduction in pressure due to Bernoulli suction is given by:
\begin{equation}
    \Delta p_1 = \dfrac{1}{2}\rho_v u_0^2-\dfrac{1}{2}\rho_v (u_0+\Delta u)^2.
\end{equation}
The change in Laplace pressure from the curvature of the interface is given by:
\begin{equation}
    \Delta p_2 \approx \gamma\dfrac{d^2h}{dx^2}=\dfrac{4\Delta h \pi^2\gamma}{\lambda^2},
\end{equation}
where the derivative is evaluated at point of closest approach. If $\Delta p_1+\Delta p_2<0$, the perturbation will grow. Combining these equations and assuming $\Delta h/h_0\ll1$, we can derive an expression for the critical wavelength that satisfies this condition:
\begin{equation}
    \lambda_c=\dfrac{2\pi\sqrt{h_0\gamma}}{u_0\sqrt{\rho_v}}.\label{lambdac}
\end{equation}

To estimate $\lambda_c$, we use the $\phi=0$ contour line (\textit{i.e.}, Fig.\ \ref{MultiplePerturbations}c) where $h_0\approx 150$ $\mu$m, $u_0\approx0.3$ m/s, $\gamma=0.0588$ N/m, and the density $\rho_v=$ 300-500 kg/m$^3$ since we do not have a sharp liquid-vapor interface (Fig.\ \ref{Unwrapped1}a). With these numerical values, $\lambda_c\approx$ 2.8-3.6 mm, which is in reasonable agreement with the wavelengths observed in Fig.\ \ref{MultiplePerturbations}b. In experiments, the liquid-vapor phase separation is much sharper and well-defined. As a consequence, to estimate $\lambda_c$, one must use the density of water vapor near boiling, $\rho_v=0.59$ kg/m$^3$. Additionally, at failure, the average thickness of the vapor layer is $h_0\approx$ 15 $\mu$m \cite{Harvey2021}. The velocity $u_0$ in experiments should be much larger than the simulations because $h_0$ is smaller, leading to enhanced heat transfer and a higher rate of evaporation, and by conservation of mass (Eq.\ \ref{masscon}). With this in mind, a mean velocity of $u_0=3.0$ m/s results in $\lambda_c=$ 2.5 mm. It is worth noting that dynamic capillary waves with wavelength 2-3 mm were observed in the experiments \cite{Harvey2021}. Thus, it is possible that failure may occur by the cooling and thinning of the vapor layer until naturally-excited waves can be suctioned to the surface by Bernoulli pressure in the vapor layer.

\begin{figure*}
    \centering
    \includegraphics[width= 6 in]{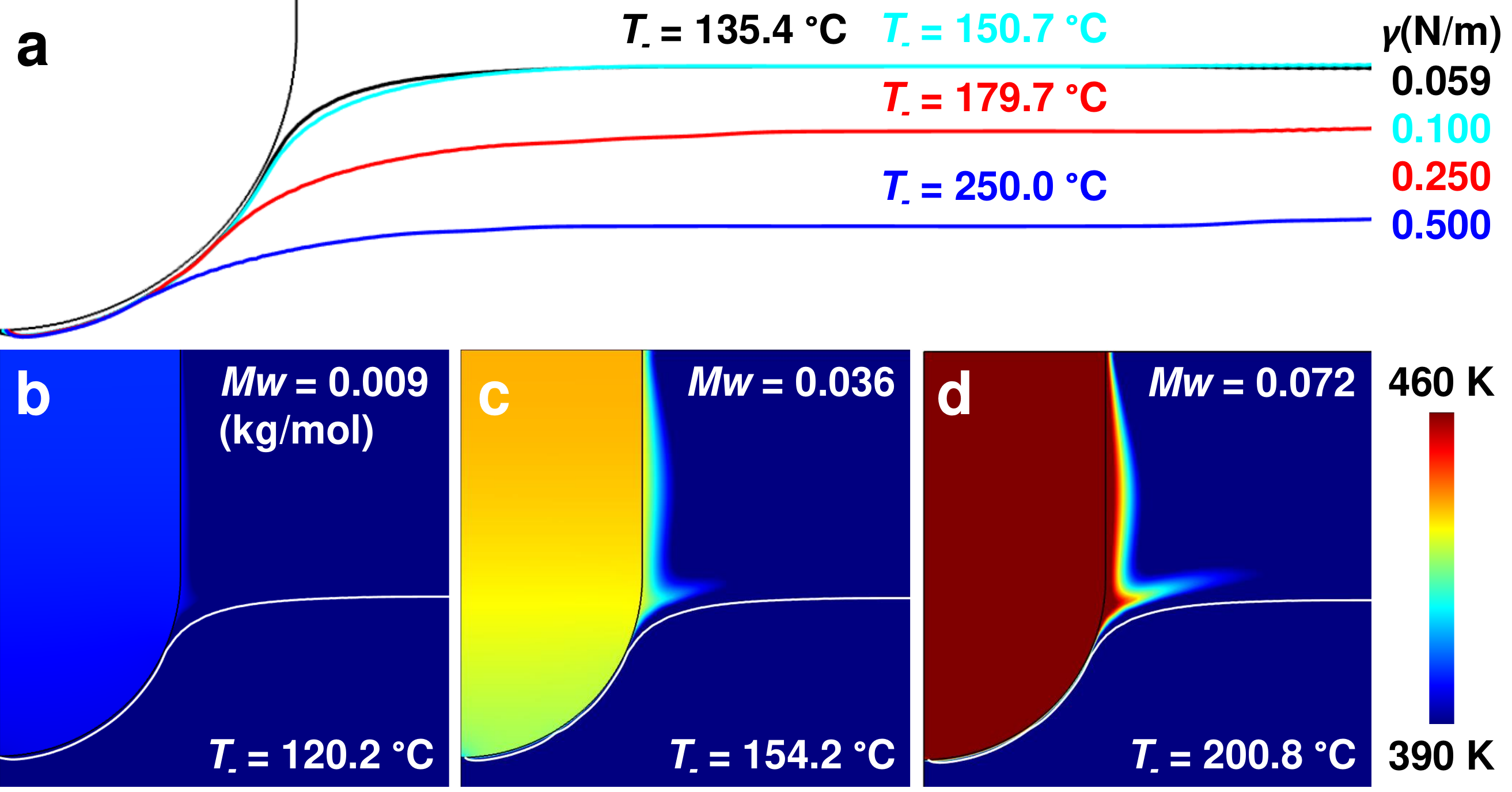}
    \caption{(a) Interface profiles just prior to failure for various values of $\gamma$ indicated in the legend. All other simulation parameters were held constant. Increasing $\gamma$ significantly changed the liquid-vapor interface profiles and increased the temperatures at failure, $T_-$. (b-c) Temperature heat maps just prior to failure when increasing the molecular weight of the gas, $Mw$. The failure temperature increases significantly while maintaining the same interface profile. All other simulation parameters were held constant.}
    \label{EffectMw}
\end{figure*}

There are 2 independent parameters in Eq. \ref{lambdac} that we can vary in our simulations: $\gamma$ and $\rho_v$. Naively, one would expect that increasing $\gamma$ would lead to lower values of $T_-$ due to its stabilizing influence against perturbations. Additionally, increasing $\rho_v$ would enhance the Bernoulli suction and lead to larger values of $T_-$. However, Fig.\ \ref{EffectMw}a shows that increasing $\gamma$ leads to \textit{larger} values of $T_-$. An important consequence of increasing the surface tension is that the liquid-vapor interface profile varies with $\gamma$. Also, the capillary length increases with $\gamma$, so that a smaller vapor layer area has to support the hydrostatic pressure of a larger surface depression. Thus, a higher temperature is required to generate enough evaporative flow to balance this pressure. Such a large increase in $u_0$ from evaporation may actually enhance the Bernoulli effect in Eq. \ref{lambdac}. 

\begin{figure}
    \centering
    \includegraphics[width=\columnwidth]{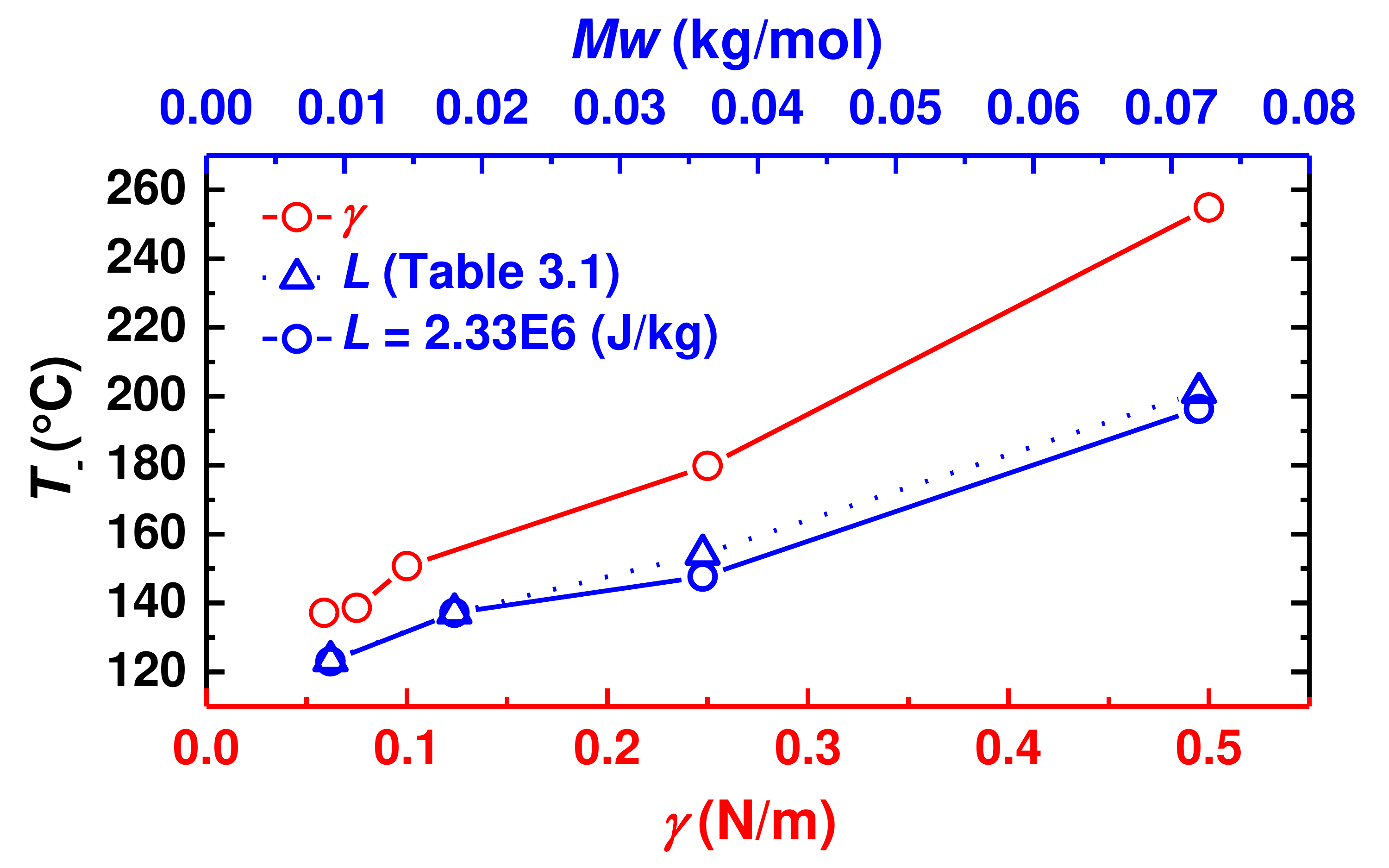}
    \caption{The resulting values of $T_-$ from simulations where either the surface tension coefficient, $\gamma$ (red), or the molecular weight of the gas, $Mw$ (blue), were varied. $Mw$ is used to define $\rho_v$ of the simulations (Table \ref{Parameters}). While varying $Mw$, the latent heat $L$ either varied with $Mw$ (triangles), or was held constant (circles), as described in the text.}
    \label{MwAndSurfaceTension}
\end{figure}

A more direct test of Eq. \ref{lambdac} can be achieved by simply changing the molecular weight of the vapor, $Mw$. The temperature-dependent vapor density, $\rho_v$, is proportional to $Mw$, as seen in Table \ref{Parameters}. Since variations in $Mw$ do not affect the surface tension or viscosity, for example, the lubrication approximation that determines the steady-state interface profile is unchanged. This is what we observe, as illustrated in Fig.\ \ref{EffectMw}b-d. Increasing $Mw$ leads to larger values of $T_-$, while the interface profile and geometry remain essentially unchanged. Examining Eq. \ref{lambdac}, increases in $\rho_v$ ($Mw$) would lead to smaller values of the critical wavelength, meaning that the interface is less stable to perturbations, and will fail at higher temperatures. Figure \ref{MwAndSurfaceTension} shows how varying $\gamma$ or $Mw$ monotonically affects $T_-$, from 120$^\circ$C-260$^\circ$C. In our simulations, we note that strictly changing $Mw$ will also change the latent heat $L$, as given in Table \ref{Parameters}. For completeness, we ran two sets of simulations, ones where we held $L$ fixed at $2.33\times10^6$ J/kg, and ones where $L$ was allowed to vary with $Mw$. Both sets showed nearly identical results for $T_-$, indicating that hydrodynamic inertial effects ($\rho_v$) played a dominant role in determining $T_-$ and the failure of the vapor layer.

\section{Conclusion}

Using simulations that combine two-phase laminar flow, heat transfer, and evaporation, we examined the formation and ultimate failure of Leidenfrost vapor layers around smooth objects. Our simulations were verified by comparing steady-state vapor layer profiles to a viscous lubrication approximation, showing excellent agreement. The failure of the vapor layer is highly dynamic and involves many thermo-physical effects not present in lubrication theory, which is often used to describe Leidenfrost vapor layers. Our results here strongly support recent experimental observations \cite{Harvey2021}: 1) $T_-$ is nearly independent of the thermal properties of the solid material, 2) $T_-$ can approach boiling temperatures for small vapor layer surface areas, and plateaus for areas larger than the capillary length of the liquid, and 3) A hydrodynamic instability determines $T_-$ and the mechanism of vapor layer failure. In the simulations, liquid-solid contact and explosive boiling occurs at a single point determined by the rapid growth of well-defined perturbations in a few milliseconds. A major finding of this work is that gas inertia, which is often completely ignored in theories of Leidenfrost vapor layers, dominates the instability leading to failure. The local reduction in pressure near a surface perturbation draws the liquid-vapor interface to the surface. These results demonstrate that the short time scales inherent in Leidenfrost vapor layers can be successfully captured in multiphysics simulations, opening new paths to study dynamic Leidenfrost phenomena. In future studies, we suggest that controllable variations in surface geometry or roughness may be harnessed to increase or decrease $T_-$, in a similar way to the formation temperature of Leidenfrost vapor layers \cite{jiang2022inhibiting}.

\begin{acknowledgments} 
We gratefully acknowledge conversations with Eric Weeks. This work was supported by the NSF DMR Grant No. 2010524. 
\end{acknowledgments} 
\appendix

\end{document}